\documentclass[12pt]{iopart}

\usepackage{amssymb}
\usepackage{cite}

\usepackage{bm}
\usepackage{upgreek}
\usepackage{graphicx}

\newcommand{\cvE}{\bm{\mathcal{E}}}
\newcommand{\cvH}{\bm{\mathcal{H}}}

\newcommand{\cv}[1]{\bm{\mathcal{#1}}}

\newcommand{\re}{\mathrm{Re}}
\newcommand{\cE}{\mathcal{E}}

\newcommand{\be}{\begin{equation}}
\newcommand{\ee}{\end{equation}}

\newcommand{\vc}[1]{\mathbf{#1}}
\newcommand{\bd}[1]{{\boldsymbol #1}}

\newcommand{\half}{{\textstyle\frac1{2}}}
\newcommand{\bn}{\bar{n}}
\newcommand{\bna}{\bn\alpha}

\newcommand{\tc}{\tilde{c}}
\newcommand{\td}{\tilde{d}}

\newcommand{\Pl}{P_l^1}
\newcommand{\Plp}{P_l^{1\prime}}
\newcommand{\Pk}{P_k^1}
\newcommand{\Pkp}{P_k^{1\prime}}

\begin{document}

\title[Laser induced interface deformation]{Theoretical considerations of laser induced liquid-liquid interface deformation}

\author{N S Aanensen$^1$, S {\AA} Ellingsen$^2$ and I Brevik$^2$}

\address{$^1$Department of Electric Power Engineering, Norwegian University of Science and Technology, N-7491 Trondheim, Norway.}
\address{$^2$Department of Energy and Process Engineering, Norwegian University of Science and Technology, N-7491 Trondheim, Norway.}
\ead{iver.h.brevik@ntnu.no}


\begin{abstract}

  In the increasingly active field of optofluidics, a series of experiments involving near-critical two-fluid interfaces have shown a number of interesting non-linear effects. We here offer, for the first time to our knowledge, an explanation for one such feature, observed in experiments by Casner and Delville  [Phys.~Rev.~ Lett. {\bf 90}, 144503 (2003)], namely the sudden formation of ``shoulder''-like shapes in a laser-induced deformation of the liquid-liquid interface at high laser power.  Two candidate explanations are the following: firstly, that the shape can be explained by balancing forces of buoyancy, laser pull and surface tension only, and that the observed change of deformation shape is the sudden jump from one solution of the strongly nonlinear governing differential equation to another. Secondly, it might be that the nontrivial shape observed could be the result of temperature gradients due to local absorptive heating of the liquid. We report that a systematic search for solutions of the governing equation in the first case yields no trace of solutions containing such features. By contrast,  an investigation of the second option  shows that the narrow shape of the tip of the deformation can be explained by a slight {\it heating} of the liquids. The local heating amounts to a few kelvins, with the parameters given, although there are  uncertainties here. Our investigations  suggest that local temperature variations are the crucial element behind the instability  and the   shoulder-like deformation.
\end{abstract}

%\today

\maketitle

\section{Introduction}\label{sec:intro}

Optical manipulation of fluid interfaces by means of lasers is a research field in rapid growth. Microfluidic applications are already diverse; general accounts can be found, for instance, in ~\cite{monat07} and \cite{monat07a}. An important advantage of the method is that it is contactless and nondestructive, and easily reconfigurable \cite{casner01,delville09,optofluidics11}. The development of this research field has taken place during several decades. Let us briefly mention three milestones of this development:

Our first example is the classic radiation experiment of Ashkin and Dziedzic \cite{ashkin73} --- cf.\ also Ashkin's  extensive reprint volume \cite{ashkin06}. Focused light was sent from above towards an air-water surface and an outward bulge of the surface of the order of 1 $\mu$m was observed. The light source was a pulsed frequency doubled Nd:YAG laser, with pulse duration 60 ns, peak power 3 kW, and beam waist 4.5 $\mu$m. The reason for the smallness of the surface elevation was the large air-water surface tension as well as the large difference in density between air and water. Theoretical treatments of this experiment can be found in ~\cite{lai76} and \cite{brevik79}.
The second example is the striking experiment of Zhang and Chang \cite{zhang88}, measuring the deformation of a micro water droplet when illuminated by a laser pulse. Typical pulse energies were 100 mJ. Theoretical papers can also be found of this effect; cf. ~\cite{lai89} and \cite{brevik99,ellingsen11}.

Third --- and that is the situation to which we will focus attention in the following --- is to decrease the fluid-fluid surface tension dramatically by working with a two-fluid system of surfactant-coated nanodroplets in oil emulsions near the critical point. A series of experimental and theoretical papers have been published  by Delville {\it et al.}   \cite{delville09, casner01b,casner03, casner03b,casner04, delville06, wunenburger06,schroll07,baroud07,brasselet08, chraibi08,chraibi10,wunenburger11} and by others \cite{hallanger05, birkeland08}. In the vicinity of the critical point the surface tension can be made about $10^6$ times smaller than the usual air-water tension and the force of gravity plays a smaller role since the difference in fluid density is small. The  displacement of the interface can accordingly be very large, about 70 $\mu$m.

In the linear regime, when laser power is low, the deformation has been satisfactorily explained theoretically using classical electrodynamics--- c.f., e.g., \cite{hallanger05} and \cite{birkeland08}. The effect is clearly illustrated, e.g.\ in   Fig.~2a of
\cite{casner03b}. The phenomenon is sketched in Fig.~\ref{fig:shoulder}a. However, when the laser power $P$ is increased, typically in excess of $600$ mW, there occurs a sudden transition into a form illustrated in Fig.~\label{fig:shoulder}b: there is produced an elongated lower channel, which we shall refer to as a protuberance. The lower channel is narrower than the upper, so that there is a small area of rapidly changing radius, a ``shoulder", in the displacement. No theory exists to our knowledge for this kind of protuberance formation. The effect  is definitely   of interest to understand, in connection with the technique of manipulating   soft matter interfaces non-invasively with radiation.

\begin{figure}[tb]
  \begin{center}
    \includegraphics[width=2.5in]{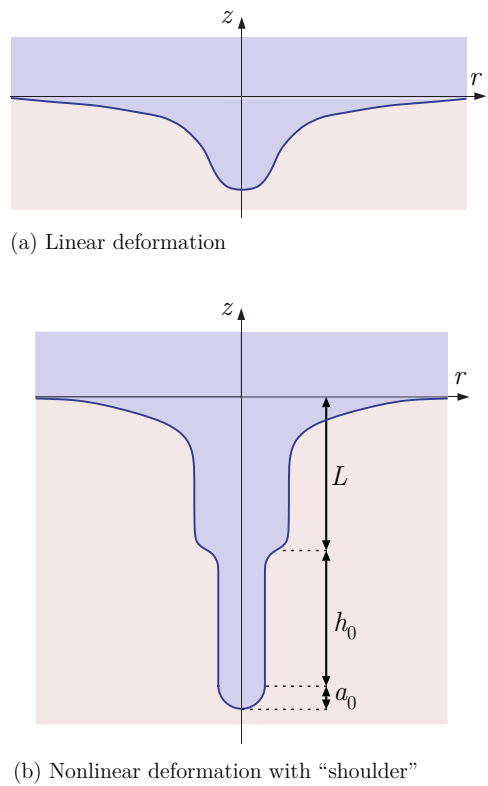}
  \end{center}
  \caption{Sketch of observed surface deformation in the linear regime, and for higher laser powers where the protuberance is observed \cite{casner03b}.}
  \label{fig:shoulder}
\end{figure}

There are  two natural possibilities  to explain the instability mechanism behind this transition:
\begin{enumerate}
\item The first possibility is that the reason is of a mathematical nature: the governing equation for the surface displacement  (Eq.~(\ref{21}) below) may contain the protuberance form as one of its solutions, even if
%the surface tension $\sigma$ is
physical parameters are
assumed to remain undisturbed by the laser beam.
\item The second possibility is of a more physical nature: the effect may be due to a local change of physical parameters $\sigma, \rho$ and $n$ caused by the high laser intensity on, and in the vicinity of, the central laser beam axis.
\end{enumerate}
In the following sections we will comment upon the first of these options, and thereafter analyze the second one in some detail. First, we will delineate some  essentials of the electromagnetic theory needed in the problem.

%%%%%%%%%%%%%%%%%%%%%%%%%%%%%%%%%%%%%%%%%%%%%%%%%%%%%%%%%%%%%%%%%%%%%%%%%%%%%%%%%%%%%%%%%%%%%%%%%%%%%%%%
\section{Basic theory}\label{sec:theory}

 Assume that laser light comes in vertically from below through medium 2, and becomes transmitted into the upper medium 1. The upper medium is the optically denser one, so that $n_1>n_2$.  We take $n_1$ and $n_2$ to be real at the actual laser frequency. As for gravity the situation is reversed so that the lower medium is the heavier one, $\rho_2>\rho_1$. The differences between material constants in the two media are in the present case small, of the order of 1\%. For the sake of clarity we take the differences to be positive quantities,
\begin{equation}
  \Delta n=n_1-n_2, \quad \Delta \rho=\rho_2-\rho_1. \label{1}
\end{equation}
The angle of incidence for the incoming wave with wave vector ${\bf k}_i$ is $\theta_i$, the angle of transmission (wave vector ${\bf k}_t$) is $\theta_t$, and the wave vector for the reflected ray is ${\bf k}_r$. The unit normal vector $\bf n$ is taken to point from medium 1 to 2\footnote{It should be mentioned that these definitions switch 1 and 2 as compared with earlier works; cf.\ \cite{hallanger05,delville06,birkeland08}.}. For numerical purposes later, we shall use numerical values for physical quantities as they appear in \cite{casner03b}.
\begin{figure}[tb]
  \begin{center}
  \includegraphics[width=3.2in]{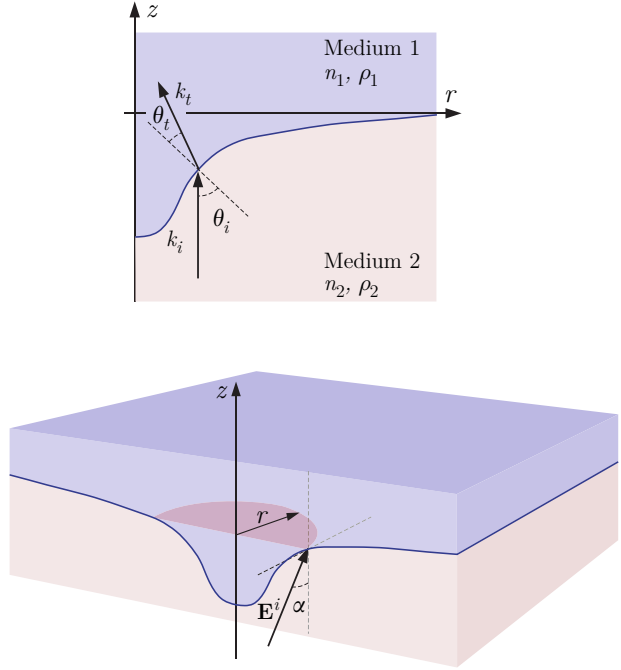}
  \end{center}
  \caption{The geometry considered: a laser impinges on a fluid-fluid interface from below, causing a downward bulge to appear. }
  \label{fig:geometry}
\end{figure}

Let us delineate how $\Delta n$, $\Delta \rho$ and the surface tension $\sigma$ vary with temperature $T$ in the vicinity of the critical temperature $T_C$.
First, according to scaling laws
\begin{equation}
  \Delta \rho=(\Delta \rho)_*\left(\frac{T-T_C}{T_C}\right)^\beta, \label{2}
\end{equation}
where $(\Delta \rho)_* = 285\,\mathrm{kg/m}^3$, with critical scaling exponent $\beta=0.325$, and $T_C = 308.15\,\mathrm{K}$ (The separation into two components of the fluid mixture occurs for $T >T_C$.).
According to the Clausius-Mossotti relation, $\Delta n\propto \Delta \rho$, so
\begin{equation}
  \Delta n=(\Delta n)_*\left(\frac{T-T_C}{T_C}\right)^\beta, \label{dn}
\end{equation}
as well, with $(\Delta n)_*=0.0321$.
As for the surface tension one has analogously
\begin{equation}
  \sigma=\sigma_*\left( \frac{T-T_C}{T_C}\right)^{2\nu}, \label{3}
\end{equation}
with $\sigma_*=1.04\times 10^{-4}$ N/m and $\nu=0.63$ (more details can be found in Ref.~\cite{casner01}).

The pressure difference across the interface $z=h$ due to surface tension is
\begin{equation}
  p_2-p_1=\sigma \left(\frac{1}{R_1}+\frac{1}{R_2}\right), \label{4}
\end{equation}
where $R_1$ and $R_2$ are the principal radii of curvature. With azimuthal symmetry, as we shall assume,
\begin{equation}
  p_2-p_1=-\frac{\sigma}{r}\frac{\mathrm{d}}{\mathrm{d}r}\left( \frac{rh_r}{\sqrt{1+h_r^2}}\right), \label{5}
\end{equation}
with $h_r=\mathrm{d}h/\mathrm{d}r$. The undisturbed surface is at $h=z=0$. According to Fig.~\ref{fig:geometry}, $h<0$ for the surface dip.

Consider now the electromagnetic surface density in the fluid (cf., for instance \cite{brevik79} or \cite{stratton41}),
\begin{equation}
{\bf f}=-\frac{1}{2}\epsilon_0\cE^2\nabla \epsilon +\frac{1}{2}\left[ \cE^2\rho \left(\frac{\partial \epsilon}{\partial \rho}\right)_T\right] +\frac{\epsilon-1}{c^2}\frac{\partial}{\partial t}(\cvE\times \cvH). \label{6}
\end{equation}
We write the constitutive relations as $ \cv{D}=\epsilon_0\epsilon \cvE$, $\cv{B}=\mu_0\cvH$, implying that $\epsilon$ is a non-dimensional quantity and that all media are assumed non-magnetic.
Calligraphic font for field quantities indicate real quantities (as opposed to complex field components which we will employ later).
The first term on the right hand side can be called  the Abraham-Minkowski term, as it is common for the Abraham and Minkowski energy-momentum tensors \cite{brevik79}. The second term is the electrostriction term, important in some experiments when the velocity of sound
in the fluid is of importance, but not in the present case where there is time enough for an elastic pressure to build up to compensate for the electromagnetic force \cite{ellingsen11b, ellingsen12b}. The last term is the so-called Abraham term, also that without importance in the present case since this term averages out over an optical period.

What is left is the Abraham-Minkowski term only, which may be called ${\bf f}^{\rm AM}$,
\begin{equation}
  {\bf f}^{\rm AM}=-\frac{\epsilon_0}{2}\cE^2 \nabla \epsilon. \label{7}
\end{equation}
By integrating this force across the boundary we obtain the vector surface force density
\begin{equation}
  {\bf \Pi}=\int_1^2{\bf f}^{\rm AM}
  \rmd
  n=\sigma^{\rm AM} \bf n, \label{8}
\end{equation}
where $\sigma^{\rm AM}$ is the scalar
\begin{equation}
  \sigma^{\rm AM}=\frac{1}{2}\epsilon_0(\epsilon_1-\epsilon_2)\left[\cE_T^2+\frac{\epsilon_1}{\epsilon_2}\cE_N^2\right]_1, \label{9}
\end{equation}
when referring to the fields on the surface in medium 1. Here ${\cvE}_T$ denotes the field component parallel to the surface, and ${\cvE}_N$ the component normal to it. If reference is instead made to the fields in medium 2,
\begin{equation}
  \sigma^{\rm AM}=\frac{1}{2}\epsilon_0(\epsilon_1-\epsilon_2)\left[\cE_T^2+\frac{\epsilon_2}{\epsilon_1}\cE_N^2\right]_2. \label{10}
\end{equation}
The surface force, in general acting in the direction of the medium of lower permittivity, is thus in the present case directed along the normal vector $\bf n$. It corresponds to $\sigma^{\rm AM}$ defined as a positive quantity.

In accordance with usual conventions we let ${\cvE}_\parallel$ denote the field component in the plane of incidence and ${\cvE}_\perp$ the component normal to it. Thus in medium 1,
\begin{equation}
  \cE_T^2=\cE_\parallel^2\cos^2 \theta_t+\cE_\perp^2, \quad \cE_N^2=\cE_\parallel^2\sin^2\theta_t, \label{11}
\end{equation}
which together with Snell's law enables us to write the surface pressure as
\begin{equation}
  \sigma^{\rm AM}=\frac{1}{2}(\epsilon_1-\epsilon_2)\left[(\cos^2 \theta_t+\sin^2\theta_i)\cE_\parallel^2+\cE_\perp^2\right]_1. \label{12}
\end{equation}
We now introduce the energy transmission coefficients $T_\parallel$ and $T_\perp$, following the conventions of Stratton \cite{stratton41}. If ${\cvE}_\parallel^{i}$ and ${\cvE}_\perp^{i}$ denote the components of the incident field in medium 1, and similar notations for the transmitted fields in medium 1, we have
\begin{eqnarray}
  T_\parallel&=&\frac{n_1}{n_2}\frac{\cos \theta_t}{\cos \theta_i}\left( \frac{\cE_\parallel^{t}}{\cE_\parallel^{i}}\right)^2= \frac{\sin 2\theta_i\sin 2\theta_t}{\sin^2(\theta_i+\theta_t)\cos^2(\theta_i-\theta_t)}, \label{13}\\
  T_\perp&=&\frac{n_1}{n_2}\frac{\cos \theta_t}{\cos \theta_i}\left(\frac{\cE_\perp^{t}}{\cE_\perp^{i}}\right)^2=\frac{\sin 2\theta_i\sin2\theta_t}{\sin^2(\theta_i+\theta_t)}, \label{14}
\end{eqnarray}
with $n_1=\sqrt{\epsilon_1}$, $n_2=\sqrt{\epsilon_2}$. Let $\alpha$ denote the angle between ${\cvE}^{(i)}$ and the plane of incidence,
\begin{equation}
  \cE_\parallel^{i}=\cE^{i}\cos \alpha, \quad \cE_\perp^{i}=\cE^{i}\sin\alpha. \label{15}
\end{equation}
Then, by introducing the mean intensity $I$ of the incoming beam,
\begin{equation}
  I=\epsilon_0n_2c\langle {\cE^{i}}^2\rangle, \label{16}
\end{equation}
we have
\begin{eqnarray}
  \sigma^{\rm AM}&=&\frac{I}{2c}\frac{n_1^2-n_2^2}{n_1}\,\frac{\cos \theta_i}{\cos \theta_t}[(\sin^2\theta_i+\cos^2\theta_t)T_\parallel  \cos^2\alpha \nonumber \\
  &&+T_\perp \sin^2\alpha]. \label{17}
\end{eqnarray}
We assume circular polarization in the following, so that $\langle \sin^2\alpha\rangle=\langle \cos^2\alpha\rangle=1/2$. It implies that  also the hydrodynamical response of the surface becomes cylindrically symmetric. Using ordinary cylindrical co-ordinates we have $\partial h/\partial \theta=0$, and $\sin\theta_i=h_r\cos\theta_i =h_r(1+h_r^2)^{-1/2}$. Let now $\bar n$ denote the relative refractive index,
\begin{equation}
  \bar{n}=\frac{n_2}{n_1}<1. \label{18}
\end{equation}
Then we can write
\begin{equation}
  \sigma^{\rm AM}(r)=\frac{2n_2I(r)}{c}\frac{1-\bar n}{1+\bar n}\, f(\bar{n}, h_r), \label{19}
\end{equation}
where $f(\bar{n}, h_r)$ is the function
\begin{eqnarray}
  f(\bar{n}, h_r) &=&\frac{1+(2-\bar{n}^2)h_r^2+h_r^4+\bar{n}h_r^2\sqrt{1+(1-\bar{n}^2)h_r^2}}{\left[\bar{n}h_r^2+\sqrt{1+(1-\bar{n}^2)h_r^2}\right]^2}\nonumber\\
  &&\times \frac{(1+\bar{n})^2}{\left[\bar{n}+\sqrt{1+(1-\bar{n}^2)h_r^2}\right]^2}. \label{20}
\end{eqnarray}
We can now write the pressure condition at equilibrium as
\begin{equation}
  (\rho_2-\rho_1)gh(r)-\frac{\sigma}{r}\frac{\mathrm{d}}{\mathrm{d}r}\left[\frac{rh_r}{\sqrt{1+h_r^2}}\right]= -\sigma^{\rm AM}(r) \label{21}
\end{equation}
(recall that since the elevation is negative, $h(r)<0$, whereas $\sigma^{\rm AM}(r)>0$ when $n_1>n_2$). The radiation pressure on the right hand side thus displaces
the surface downward, while the surface tension term acts upward, opposing the depression. The sum of these two effects must be balanced by the influence of gravity (i.e., buoyancy) for mechanical equilibrium.
Equation (\ref{21}) is our governing equation. Note that in this section the surface tension $\sigma$ has been assumed constant.

%%%%%%%%%%%%%%%%%%%%%%%%%%%%%%%%%%%%%%%%
%%%%%%%%%%% S E C T I O N %%%%%%%%%%%%%%
%%%%%%%%%%%%%%%%%%%%%%%%%%%%%%%%%%%%%%%%
\section{First option: Investigation of a set of trial functions}\label{sec:numerics}

It is quite natural to check if the governing equation (\ref{21}) possesses solutions corresponding  to an abrupt jump from one kind of surface deflection to another, completely different one.  If there are special solutions compatible with the protuberance form seen in the  experiments it should be possible to retrace their form, at least approximately, by  inserting reasonable test solutions into the equation itself. If we are on the right track, the difference between the right and left hand sides of the governing equation should be small.

A number of trial functions were used, guided by experimentally observed shapes. Equations were put into non-dimensional form, introducing the Bond number  $B_0$ and the capillary length $l_c$ as
\begin{equation}
  B_0 = \left(\frac{w_0}{l_c}\right)^2, ~~~l_c = \sqrt{\frac{\sigma}{(\rho_2-\rho_1)g}} \label{22},
\end{equation}
with  $w_0$  the laser beam waist. Convenient non-dimensional (positive) variables for the height and radius of the deformation were
\begin{equation}
  H = -\frac{h}{l_c},   ~~~  R(H) = \frac{r(h)}{w_0}. \label{23}
\end{equation}
Assuming a Gaussian laser beam, the intensity could be written in terms of the laser power, $P$, as
\be
  I(r) = \frac{2P}{\pi w_0^2} \rme^{-2R^2}. \label{24}
\ee
The governing equation could thus be expressed in non-dimensional form, with boundary conditions
\begin{equation}
  \lim_{H\to 0}R = \infty,
  ~~~
  \lim_{H\to 0} R_H = -\infty.\label{26}
\end{equation}
We do not go into any detail concerning the use of these trial expressions; the main conclusion that we can make is that it is highly unlikely that there are special solutions of Eq.~(\ref{21}) compatible with the observed form. This is consonant with the finding of Refs.~\cite{chraibi08,chraibi10}, where a direct numerical simulation method is used.
%SAE changed:
In no case did the introduction of a ``shoulder'' shape improve the solution (i.e.\ tend to better equate the left and right hand side of Eq.~(\ref{21})); always the contrary was observed. Although the result was negative, this brief discussion might be a useful inclusion to the future researcher of this question.

%%%%%%%%%%%%%%%%%%%%%%%%%%%%%%%%%%%%%%%%
%%%%%%%%%%% S E C T I O N %%%%%%%%%%%%%%
%%%%%%%%%%%%%%%%%%%%%%%%%%%%%%%%%%%%%%%%
\section{Second option: Local variations of physical parameters. Heating effects}\label{sec:thermal}

Now focus attention on the physics of the situation: the central region of the laser beam where the field is strongest, may be expected to heat
 the fluids in the central region causing the physical parameters $\Delta n, \Delta \rho$ and $\sigma$ to attain locally different values. A different shape can accordingly in principle be the result of the condition of  mechanical equilibrium establishing
a cylinder-like lower protuberance of radius $a_0$ and length $h_0$;
cf.\ Fig.~\ref{fig:shoulder}.
The dimensions and other characteristics of the protuberance are determined by this equilibrium condition.

In this section we shall use the values of $h_0$ and $a_0$ measured in experiment, together with the known physical parameters at temperature $310.65$K and the temperature behaviour of $\Delta n$ and $\sigma_0$ to estimate what local temperature near the tip might give rise to a protuberance of this smaller radius.
The assumption made is thus that the heating by the laser is restricted to a small area near the tip of the deformation. We do not, therefore, include in our calculation the change to the density of the fluid in the boyancy term of the force balance, since that term is an integral over the full volume of the protuberance, of which only a small charge will be affected. The temperature change does, however, tend to increase the local surface tension $\sigma(T)$ and the difference in index of refraction, $\Delta n$. The former increase tends to pull the protuberance upward (smaller value of $h_0$) whereas the latter increases the electromagnetic radiation force and tends to pull downwards.

Let us go through the details of the calculation.
The mechanical forces acting on the protuberance region are
as before from three different contributions: buoyancy (i.e.\ gravity or hydrostatic pressure), surface tension and radiation force.
The pressure at its upper end is the hydrostatic pressure $\rho_1 gL$, where $L$ denotes the height of the depression at the laser power just before the protuberance becomes formed (cf. Fig. 1). Similarly, the pressure of the outside of the lower tip is also the hydrostatic pressure. At the lower end of the cylindrical section it is  $\rho_2 g(L+h_0)$ (for simplicity we take all heights to be positive quantities). The buoyancy force on the protuberance is accordingly, when we include the volume of the hemisphere,
\begin{equation}
  B =F_B=g\pi a_0^2\left[ (L+h_0+\frac{2}{3}a_0)\Delta \rho+\rho_1(h_0+\frac{2}{3}a_0)\right], \label{100}
\end{equation}
where we have used $\rho_2=\rho_1+\Delta \rho$. Subtracting off the weight $W=g\pi a_0^2\rho_1(h_0+\frac{2}{3}a_0)$ of the liquid in the protuberance region we get the  upward directed
buoyancy
force
\begin{equation}
F_B-W=g\pi a_0^2(L+h_0+\frac{2}{3}a_0)\Delta \rho. \label{101}
\end{equation}
Although the simplified test geometry thus implicitly assumed underestimates the volume of the deformation, we find that the results obtained are very insensitive to this. Increasing the volume by $50\%$, for example, only changes the estimated temperature increase by about $0.2$K, which is not significant given the simplicity of the model itself.

\begin{figure}[tb]
  \begin{center}
    \includegraphics[width=2.4in]{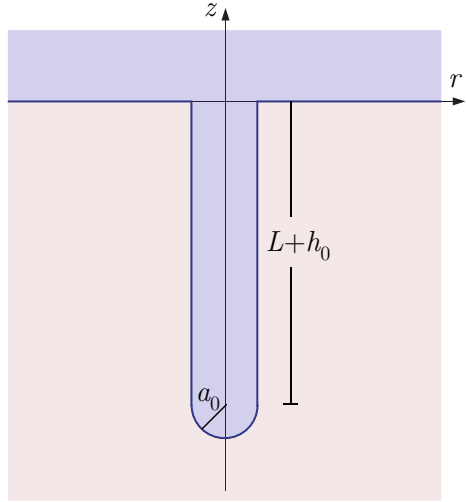}
  \end{center}
  \caption{Model geometry for estimation of tip temperature.}
  \label{hemisphere}
\end{figure}

The buoyancy force has to be supplemented  with the surface tension force $2\pi \sigma_0a_0$, also acting upwards. The third force, acting downwards, is the radiation force $F_{\rm rad}$ on the lower hemisphere tip. We shall work with the magnitude $|F_{\rm rad}|$ of $F_{\rm rad}$ to avoid negative quantities. The condition for mechanical equilibrium of the protuberance can now be written as
 \begin{equation}
g\pi a_0^2(L+h_0+\frac{2}{3}a_0)\Delta \rho + 2\pi \sigma_0a_0=|F_{\rm rad}|. \label{102}
\end{equation}
Once $F_{\rm rad}$ is known - cf. the next section -  Eq.~(\ref{102}) determines  the lower-tip surface tension $\sigma_0$. Recall that the input parameters, to be inferred from experiments, are   $\Delta \rho, L, h_0$, and $a_0$.

We introduce now a quantity $Q$, as a non-dimensional measure of the
force exerted by
the incident laser beam. It is defined as follows: If the beam were plane, with intensity $I_0$, we would according to common usage write the radiation force $F_{\rm rad}$ (here taken positive) on a cross-sectional area $\pi a_0^2$ as
\begin{equation}
F_{\rm rad}=\frac{I_0n_2}{c}\pi a_0^2 Q, \label{103}
\end{equation}
meaning that a perfect absorber corresponds to $Q=1$. Considering instead a Gaussian beam with total power $P$ and beam waist $w_0$ we have, if $I_0$ now means the intensity on the symmetry axis,
\begin{equation}
I_0=\frac{2n_2P}{\pi w_0^2}. \label{104}
\end{equation}
Accordingly, we can in the present case write
\begin{equation}
F_{\rm rad}=\frac{2n_2^2Pa_0^2}{w_0^2c}Q. \label{105}
\end{equation}

%%%%%%%%%%%%%%%%%%%%%%%%%%%%%%%%%%%%%%%%
%%%%%%%%%%% S E C T I O N %%%%%%%%%%%%%%
%%%%%%%%%%%%%%%%%%%%%%%%%%%%%%%%%%%%%%%%
\subsection{Electromagnetic force on a hemisphere}

As is known, the force acting on a closed surface $\mathcal{S}$ may be calculated by integrating the stress tensor
\be
  \bd{T}=\cvE\otimes\cv{D}+\cv{H}\otimes\cv{B}-\half(\cv{E}\cdot\cv{D}+\cv{H}\cdot\cv{B})\bd{1}
\ee
over the surface
\be
  F_\mathrm{rad} = \oint_\mathcal{S}dS \vc{\hat r}\cdot \langle \bd{T} \rangle \cdot \vc{\hat z}=\oint_\mathcal{S}dS \vc{\hat r}\cdot \langle T_{rz} \rangle \cdot \vc{\hat z}
\ee
with $\bd{1}$ the unit matrix and $\langle \cdots\rangle$ denotes time average and hats denote unit vectors. (For an isotropic, dielectric medium the tensors of Minkowski and Abraham coincide. The Abraham force \cite{brevik79} oscillates out and gives no contribution in the optical case.) Here
\be
  \cv{E} = \re\{\vc{E} \rme^{\rmi\omega t}\},~~ \mbox{etc.,}
\ee
where $\vc{E},\vc{D},\vc{B}$ and $\vc{H}$ are complex field vectors. For two field components $\bar{X}_i,\bar{X}_j$ we have $\langle \bar{X}_i\bar{X}_j\rangle = \half \re\{
X_iX_j^*\}$.

In the case of a full sphere, or indeed the total force on any isolated body in a homogeneous external medium, it may be opportune to integrate the stress tensor over a closed surface far from the body in order to make use of simplified asymptotic expressions for the spherical Bessel functions involved. When evaluating the force on a part of the body surface, however, integration must be performed at the actual interface, and we find it simpler in this case to
express the stress tensor both outside and inside the spherical surface in terms of the \emph{interior} fields, since the external fields have both incident a scattered components. Using the continuity of $D_r,E_\theta,E_\phi$ and $\vc{H}$ across the surface,  the axial front force on a sphere segment $\theta_0 < \theta < \pi, 0<\phi <2\pi$ can be expressed as
\begin{eqnarray}
  F_\mathrm{rad} &=& a_0^2\int_0^{2\pi}\rmd \phi \int_{\theta_0}^\pi\sin\theta \rmd \theta\langle T_{rz}^\mathrm{ext}-T_{rz}^\mathrm{int}\rangle\nonumber\\
  &=& \frac{\pi n_2^2a_0^2\varepsilon_0}{2} (\bn^2-1)\int_{\theta_0}^\pi(\bn^2|E^w_r|^2+|E^w_t|^2)\nonumber \\
  &&\times\cos\theta\sin\theta \rmd \theta
\end{eqnarray}
where $|E_t|^2 = |E_\theta|^2+|E_\phi|^2$, and we have used the fact that the integrand has no $\phi$ dependence. Superscript $w$ signifies that the fields are evaluated just within the surface of the hemisphere, at radius $r=a^-$. The expression is obviously real and there is no longer a need to explicitly take the real part.

We now make the assumption that the electromagnetic fields inside the hemispherical surface equal those inside a full sphere, and use the electromagnetic field expressions due to Barton and co-workers \cite{barton89}, which we quote in some detail in the Appendix. As before, we let the incident field be a circularly polarized plane wave, for which the explicit field expansions are found in Eqs.~(\ref{CircPolFieldsr})-(\ref{CircPolFieldsph}), and we assume the laser width $w_0$ to be sufficiently large that laser intensity can be approximated as uniform over the hemisphere (this is a reasonable approximation for our purposes since the optical force on a hemisphere is quite concentrated near the symmetry axis \cite{ellingsen11}.).

We make the convenient definitions
%\begin{subequations}
\begin{eqnarray}
  c_{l} &=& \rmi^{l+2}[\bn\psi_l(\bna)\xi_l^{(1)\prime}(\alpha)-\psi'_l(\bna)\xi_l^{(1)}]^{-1}\label{cldla}\\
  d_{l} &=& \rmi^{l+1}[\psi_l(\bna)\xi_l^{(1)\prime}(\alpha)-\bn\psi'_l(\bna)\xi_l^{(1)}]^{-1}\label{cldlb}
\end{eqnarray}
%\end{subequations}
where $\psi_l$ and $\xi_l$ are Riccati-Bessel functions of order $l$,
\be
  \alpha = kr = \frac{2\pi a_0}{\lambda_2}=\frac{n_2\omega a_0}{c}
\ee
is the number of hemisphere circumferences per optical wavelength (in medium 2).
Thus we find the radiation force on the sphere segment to be
\begin{eqnarray}
&  Q=\frac{\bn^2-1}{2\alpha^4} \sum_{k=1}^\infty\sum_{l=1}^\infty\Bigl\{ c_kc_l^*\psi_k(\bna)\psi_l(\bna)I_{kl}\nonumber\\
  &+\alpha^2[c_kc_l^*\psi'_k(\bna)\psi'_l(\bna)+d_kd_l^*\psi_k(\bna)\psi_l(\bna)]M_{kl}\nonumber\\
  &+\alpha^2[c_kd_l^*\psi'_k(\bna)\psi_l(\bna)+d_kc_l^*\psi_k(\bna)\psi'_l(\bna)]N_{kl}\Bigr\} \label{Frad}
\end{eqnarray}
where the coefficients $I_{kl}, M_{kl}$ and $N_{kl}$ contain the actual polar angle integration over $\theta$ from $\pi/2$ to $\pi$ and are given as
%\begin{subequations}
\begin{eqnarray}
  I_{kl}&=&(2k+1)(2l+1)\int_{-1}^0\rmd uu \Pk(u)\Pl(u),\\
  M_{kl}&=&\frac{2k+1}{k(k+1)}\frac{2l+1}{l(l+1)}\int_{-1}^0\frac{\rmd uu}{1-u^2}[\Pk(u)\Pl(u)\nonumber \\
  &&+(1-u^2)^2\Pkp(u)\Plp(u)],\\
  N_{kl}&=&-\frac{2k+1}{k(k+1)}\frac{2l+1}{l(l+1)}\int_{-1}^0\rmd u\Pk(u)\Pl(u).
\end{eqnarray}
%\end{subequations}
Numerically, these coefficients may be tabulated once and for all, and calculation is thus not particularly expensive. One finds that the sums in (\ref{Frad}) can be truncated at a value somewhat greater than $\alpha$. In our case $\alpha$ is in the order of $60$, so about $4000$ terms were evaluated to calculate $Q$.

%%%%%%%%%%%%%%%%%%%%%%%%%%%%%%%%%%%%%%%%
%%%%%%%%%%% S E C T I O N %%%%%%%%%%%%%%
%%%%%%%%%%%%%%%%%%%%%%%%%%%%%%%%%%%%%%%%
\subsection{Numerical results}

Given the value of $Q$ calculated from (\ref{Frad}) using the unperturbed $\Delta n$, using the fact that $Q\propto \Delta n(T)$ and inserting the temperature dependences for $\Delta n, \Delta \rho$ and $\sigma_0$ from Eqs.~(\ref{2})-(\ref{3}), temperature $T$ is the only unknown in the equation of mechanical equilibrium, and may thus be determined from Eq.~(\ref{102}). Explicitly we write
\be
  B(T_0) + S(T_0)\Theta^{2\nu} - |F_\mathrm{rad}|(T_0)\Theta^{\beta}=0
\ee
with $B$ the buoyancy force, $S$ the contribution from surface tension [see eq.~(\ref{102})], and
\be
  \Theta = \frac{T-T_C}{T_0-T_C}; ~~~ T = T_C + \Theta (T_0-T_C)
\ee
with $T_0=310.65$, the external temperature.
The equation is easily solved with respect to $\Theta$ using Newton's method.

Table \ref{tab:newtonT} shows  values of $\sigma_0$ and $\Delta n$, and corresponding tip temperatures $T_{\rm new}$, in
the cases for which experiments were performed \cite{casner01,casner03b,delville06}. There are several conclusions to be drawn from these values:
\begin{enumerate}
  \item It is apparent that $\sigma_0$ is increased in comparison with the initial global value $\sigma=2.413\times 10^{-7}$ N/m,  calculated from Eq.~(\ref{3}).
  \item Correspondingly, the new tip temperature  $T_{\rm new}$ is also increased. There occurs thus a {\it heating} of the tip, in accordance with the expectation. It should here be noted that the relationship between surface tension and temperature is in our case counterintuitive: increasing surface tension means increasing temperature, in the region above the critical point.
  \item\label{Tdiff} The increase in temperature is relatively large; for moderate powers $P$  the values of  $T_{\rm new}$ lie about 4 degrees higher than the ambient temperature $T_0$. It may be of interest to compare this with the much smaller temperature increase occuring in a homogeneous fluid illuminated by a laser beam. Adopting the Gaussian form for $I(r)$  from Eq.~(\ref{24}) we may solve the heat conduction equation
  \begin{equation}
    \nabla^2 T(r)+\frac{\alpha}{\kappa} I(r)=0 \label{X}
  \end{equation}
  in cylindrical symmetry, where $\alpha \approx 3\times 10^{-4}$ cm$^{-1}$ is the thermal absorption coefficient and $\kappa=1.28 \times 10^{-3}$ W cm$^{-1}$ K$^{-1}$ is the thermal conductivity. One finds \cite{casner01,chraibi08} that on the symmetry axis $r=0$ the local temperature increase is $\Delta T  \approx \alpha P/(4\pi \kappa)(\ln 100+ \gamma)$, where $\gamma\approx0.5772$ is Euler's constant. With $P=1$ W this amounts to about 0.1 K, which is a very moderate heating.
\end{enumerate}

The difference in expected temperature increase mentioned in point (\ref{Tdiff}) is not as unreasonable as it seems. Our calculation concerns a \emph{local} effect, the heating of the interface near the protuberance tip and its immediate surroundings, whereas the calculation in point (\ref{Tdiff}) is a heating of a much larger volume of fluid. To heat a finite body of fluid by the same extent, would require a much higher power. A significantly larger heating is expected near the tip of the protuberance anyway, since the deformation acts as a lens, focussing the incoming light and giving rise to local intensity maxima much higher than that near a laser beam in a homogeneous medium (see e.g.\ \cite{ellingsen12b}).

To further illustrate the importance of local geometry, consider the following simple argument: Let the interface be modeled as a thin horizontal plate of thickness $\Delta z$ and surface area $A=\pi a^2$, surrounded by a vacuum, illuminated by a laser power $P$. When thermal equilibrium is established, the rate of absorbed heat $\alpha P$ has to balance the rate of radiated energy to both sides, i.e. $2A\sigma_{SB}(T^4-T_0^4)$, where $\sigma_{SB}$ is the Stefan-Boltzmann constant. Since the excess temperature $\Delta T=T-T_0$ is small, $\Delta T/T_0 \ll 1$, we get the relationship
\begin{equation}
\Delta T=\frac{\alpha P}{8\pi \sigma_{SB}T_0^3} \frac{\Delta z}{a^2}. \label{Y}
\end{equation}
Inserting $\alpha$ as above, and with $P=1$ W, $\sigma_\mathrm{SB}=5.67\times 10^{-8}$\,Wm$^{-2}$K$^{-4}$, $T_0=300$ K, we get
\begin{equation}
\Delta T=(7.80\times 10^{-4}\, \mathrm{Km}) \frac{\Delta z}{a^2}. \label{Z}
\end{equation}
Now take for definiteness $a=4~\mu$m. If we choose the thickness to be small, $\Delta z=1$ nm, we get only a small temperature increase, $\Delta T \approx 0.05$ K. With $\Delta z=100$ nm the result is much higher, $\Delta T \approx 5$ K.  We thus see that the thermal behaviour is highly sensitive to the thickness over which absorption takes place. In other words, local temperature variations can take very different values from from overall (global) ones.

\begin{table}[tb]
\begin{centering}
\begin{tabular}{ccccc|ccc}
\hline
	$w_0$ [$\mu$m]&	$P$ [mW]&	$L + h_0$ [$\mu$m]&	$a_0$ [$\mu$m]&	Q [$10^{-3}$]&$\Delta n$&$\sigma_0$ [$10^{-7}$N/m]	&	$T_{\rm new}$[K]\\ \hline%\hline
	6.3&	1200&	70&	3.2&		4.564&		0.0109&	15.7		&319.1\\ 	
	6.3&	600&	40&	3.8&		4.562&		0.0090&	7.57		&314.3 \\
	4.8&	1200&	72&	2.5&		4.567&		0.0121&	24.0		&323.6 \\
	4.8&	600&	43&	2.6&		4.567&		0.0096&	9.80		&315.7 \\ \hline
\end{tabular}
\caption{Parameter sets from experiments \cite{casner01,casner03b,delville06} and calculated value of non-dimensional radiation force $Q$ from these. The requirement of mechanical equilibrium allows us to estimate the new tip temperature $T_\mathrm{new}$ and locally diffent values of $\Delta n$ and $\sigma_0$. Parameter values are $g=9.81$m/s$^2, \Delta \rho = 59.61$kg/m$^3, T_0=310.65$K$, \sigma(T_0)= 2.413\times 10^{-7}$N/m, $n_2=1.46, \Delta n(T_0)=0.00672$.}
\label{tab:newtonT}
\end{centering}
\end{table}

\section{Concluding remarks}

We thus arrive at an explanation of the observed effect that seems quite plausible, namely that the tip is heated locally and so makes the interface thermodynamically nonuniform.
The change in physical parameters $\Delta n$ and $\sigma_0$ conspire to favour a narrower deformation. We have performed a fairly simple model calculation which indicates that this is the case, yet does not constitute a full explanation for the particular shape that the deformation takes. 

It seems likely, however, that the increase in surface tension in the areas where the most laser light is absorbed, i.e.\ at the tip of the emulsion and at the ``shoulders'', could relate directly to the onset of a Plateau--Rayleigh-like instability\cite{eggers97}. A Plateau-Rayleigh instability is driven by surface tension trying to decrease the interface area, and surface tension is increased whereever the liquids are heated. As we have discussed above, a deformed surface will act as a lens, causing local light intensities to far exceed average ones near areas of sharp deformations, such as ``shoulders'' and tip. A slight change in shape increases the heating due to light focussing, increasing the surface tension and giving further change in shape, and so on. In this way, local heating can be thought to drive the required instability.
Instead of a separate droplet forming (as indeed it does for even higher laser powers \cite{casner03}), the radiation pressure may serve the role of stabilizing the intermediate ``shoulder'' geometry, similar to the situation reported in \cite{casner04}. A further investigation into this stability issue is certainly warranted, and of great potential interest.

The other option considered in this paper, namely that the shape can be explained merely by  balancing the forces of laser pull, buoyancy and surface tension with unchanged physical parameters, turned out not to be supported. Our extensive numerical search in this direction led to no indication that the unperturbed force balance equation has solutions at all similar to the observed deformations. Of course, trials of this sort cannot lead to a decisive falsification. Nevertheless,
we feel our investigation supports a firm conviction that
that the observed deformations have their roots in local effects connected with temperature variations, and
cannot be due to
global effects.

 Finally, we suggest that the  proposed explanation should be possible to confirm experimentally, were the
 experiment  \cite{casner03} to be set up again. We envisage that use of an infrared (thermal) camera could enable detection of local temperature gradients. A more complete theory could moreover be assisted by a simulation in which absorptive heating were taken into account, although this would be a project suited for heavy numerics rather than analytical means. In light of the large and growing interest in fluid manipulation with laser light, however, understanding the instability studied herein could be of some importance.

%%%%%%%%%%%%%%%%%%%%%%%%%%%%%%%%%%%%%%%%%%%%%%%%%%%%%%%%%%%%%%%%%%%%%%%%%%%%%%%%%%%%%%%%%%%%%%%%%%%%%%%%

%%%%%%%%%%%%%%%%%%%%%%%%%%%%%%%%%%%%%%%%%%%%%%%%%%%%%%%%%%%%%%%%%%%%%%%%%%%%%%%%%%%%%%%%%%%%%%%%%%%%%%%%

\appendix

\section{General formulae}

\subsection{From Mie theory}

The internal electric field components are expanded in spherical harmonics according to \cite{barton89}
%\begin{subequations}\label{generalFields}
\begin{eqnarray}
  E_r^w&=& E_0 \sum_{l=1}^\infty\sum_{m=-l}^ll(l+1)\tc_{lm}\psi_l(\bna)Y_{lm}(\Omega)\label{generalFieldsa}\\
  E_\theta^w&=& \alpha E_0\sum_{l=1}^\infty\sum_{m=-l}^l \Bigl[\bn \tc_{lm}\psi'_l(\bna)\partial_\theta Y_{lm}(\Omega)\nonumber \\
  &&-\frac{\td_{lm}}{n_2}m\psi_l(\bna)\frac{Y_{lm}(\Omega)}{\sin\theta}\Bigr]\label{generalFieldsb}\\
  E_\phi^w&=& \rmi\alpha E_0\sum_{l=1}^\infty\sum_{m=-l}^l \Bigl[m\bn \tc_{lm}\psi'_l(\bna)\frac{Y_{lm}(\Omega)}{\sin\theta}\nonumber \\
  &&-\frac{\td_{lm}}{n_2}\psi_l(\bna)\partial_\theta Y_{lm}(\Omega)\label{generalFieldsc}
\end{eqnarray}
%\end{subequations}
where $\Omega=\theta,\phi$ and the coefficients
%\begin{subequations}
\begin{eqnarray}
  \tc_{lm} &=& \rmi A_{lm}[\bn^2\psi_l(\bna)\xi_l^{(1)\prime}(\alpha)-\bn\psi'_l(\bna)\xi_l^{(1)}]^{-1}\\
  \td_{lm} &=& \rmi B_{lm}[\psi_l(\bna)\xi_l^{(1)\prime}(\alpha)-\bn\psi'_l(\bna)\xi_l^{(1)}]^{-1}
\end{eqnarray}
%\end{subequations}
and the incident field $\mathbf{E}^i$ is contained in the quantities
\be\label{AB}
  \begin{array}{c}A_{lm}\\B_{lm}\end{array} = \frac1{l(l+1)\psi_l(\alpha)}\int\left.\begin{array}{c}E_r^i/E_0\\H_r^i/H_0\end{array}\right.Y_{lm}^*(\Omega)\rmd \Omega,
\ee
where incident fields are evaluated at $r=a^+$ and the integral is over all solid angles. Here $H_0=E_0/(c\mu_0)$.

\subsection{Circularly polarised plane wave}

A circularly polarised plane wave propagating along the $z$ direction may be expressed as (\cite{jackson}, section 10.3)
\begin{eqnarray}
  \vc{E}^i&=&E_0\bd{\Lambda} \rme^{\rmi kz} =E_0\sum_{l=1}^\infty \rmi^l\sqrt{4\pi(2l+1)}[j_l(kr)\bd{X}_{l1}\nonumber\\
  &&+\frac1{k}\nabla\times j_l(kr)\bd{X}_{l1}]
\end{eqnarray}
with
\be
  \vc{X}_{lm}(\Omega)=\frac1{\rmi\sqrt{l(l+1)}}(\vc{r}\times\nabla) Y_{lm}(\Omega)
\ee
where $k=n_2\omega/c$ and $\bd{\Lambda}=\vc{\hat{x}}+\rmi\vc{\hat{y}}$. The radial component may then be written (\cite{jackson}, section 10.4) at $r=a^+$
\be\label{Eir}
  E^i_r= \frac{E_0}{\alpha^2}\sum_{l=1}^\infty \rmi^{l+1}\sqrt{4\pi l(l+1)(2l+1)}\psi_l(\alpha) Y_{l1}(\Omega).
\ee
The radial magnetic component is now found from Maxwell's equations as
\be
  H^i_r=-\frac{\rmi n_2}{\mu_0 c}E^i_r.
\ee
By using formulas (\ref{AB}) and (\ref{Eir}) we find
\be
  A_{lm} = \frac{\rmi^{l+1}}{\alpha^2}\sqrt{\frac{4\pi(2l+1)}{l(l+1)}}\delta_{m1}=\frac{\rmi}{n_2}B_{lm}
\ee
and the field components (\ref{generalFieldsa}), (\ref{generalFieldsb}) and (\ref{generalFieldsc}) may be written
%\begin{subequations}\label{CircPolFields}
\begin{eqnarray}
  E_r^w=&\frac{ \rme^{\rmi\phi}E_0}{\bn \alpha^2}\sum_{l=1}^\infty (2l+1)c_l\psi_l(\bna)P_l^1\label{CircPolFieldsr}\\
  E_\theta^w=&-\frac{ \rme^{\rmi\phi}E_0}{\alpha}\sum_{l=1}^\infty\frac{2l+1}{l(l+1)}\Bigl[c_l\psi'_l(\bna)\Plp\sin\theta\nonumber \\
  &  +d_l\psi_l(\bna)\frac{\Pl}{\sin\theta}\Bigr]\label{CircPolFieldsth}\\
  E_\phi^w=&\frac{\rmi \rme^{\rmi\phi}E_0}{\alpha}\sum_{l=1}^\infty\frac{2l+1}{l(l+1)}\Bigl[c_l\psi'_l(\bna)\frac{\Pl}{\sin\theta}\nonumber \\
    &  +d_l\psi_l(\bna)\Plp\sin\theta\Bigr]\label{CircPolFieldsph}
\end{eqnarray}
%\end{subequations}
having suppressed the argument $\cos\theta$ of the Legendre polynomials. Coefficients $c_l$ and $d_l$ are defined in Eqs.~(\ref{cldla}) and (\ref{cldlb}).


\section*{References}

\begin{thebibliography}{99}
\bibitem{monat07}
Monat C, Demachuk P, Grillet C, Collins M, Eggleton B F, Cronin-Golomb M, Mutzenich S, Mahmud T, Rosengarten G and Mitchell A 2007, {\it Microfluidics} {\bf 4} 81.
\bibitem{monat07a}
Monat C, Domachuk P, and Eggleton B J 2007,  {\it Nature Photonics} {\bf 1} 106.
\bibitem{casner01}
Casner A 2001 {\it Ph.D.\ dissertation} Universit\'e Bordeaux I, Bordeaux, France; ({\it Online:} http://tel.ccsd.cnrs.fr/documents/archives0/ 00/00/16/37/index.html).
\bibitem{delville09}
Delville J-P, de Sait Vincent M R, Schroll R D, Chra\"{\i}bi H, Issenmann B, Wunenburger R, Lasseux D, Zhang W W and Brasselet E 2009 {\it J. Opt. A} {\bf 11} 034015.
\bibitem{optofluidics11}
 {\it Optofluidics Surging Forward} 2011 focus issue of {\it Nature Photonics} {\bf 5} issue 10.
\bibitem{ashkin73}
Ashkin A and Dziedzic J M 1973  {\it Phys. Rev. Lett.} {\bf 30} 139.
\bibitem{ashkin06}
Ashkin A 2008 {\it Optical Trapping and Manipulation of Neutral Particles  Using Lasers: a Reprint Volume with Commentaries} (Singapore: World Scientific).
\bibitem{lai76}
Lai H-M and Young K 1976 {\it Phys. Rev. A} {\bf 14} 2329.
\bibitem{brevik79}
Brevik I 1979 {\it Phys. Rep.} {\bf 52} 133.
\bibitem{zhang88}
Zhang J-Z and Chang R K 1988 {\it Optics Letters} {\bf 13} 916.
\bibitem{lai89}
Lai H-M, Leung P T, Poon K L and Young K 1989 {\it J. Opt. Soc. Am. B} {\bf 6} 2430.
\bibitem{brevik99}
Brevik I and Kluge R 1999 {\it J. Opt. Soc. Am. B} {\bf 16} 976.
\bibitem{ellingsen11}
Ellingsen S {\AA} 2012 {\it Phys. Fluids} {\bf 24} 022002.
\bibitem{casner01b}
Casner A and Delville J-P 2001 {\it Phys. Rev. Lett.} {\bf 87} 054503.
\bibitem{casner03}
Casner A and Delville J-P 2003 {\it Phys. Rev. Lett.} {\bf 90} 144503.
\bibitem{casner03b}
Casner A, Delville J-P and Brevik I 2003 {\it J. Opt. Soc. Am. B} {\bf 20} 2355.
\bibitem{casner04}
Casner A and Delville J-P 2004 EPL {\bf 65} 337.
\bibitem{delville06}
Delville J-P, Casner A, Wunenburger R and Brevik I 2006 Optical deformability of fluid interfaces {\it Trends in Laser and Electro-Optics Research} ed Arkin W T (New York: Nova Science Publishers) p 1 ({\it Preprint} physics/0407008).
\bibitem{wunenburger06}
Wunenburger R, Casner A and Delville J-P 2006 {\it Phys. Rev. E} {\bf 73} 036314; {\bf 73} 036315.
\bibitem{schroll07}
Schroll R D, Wunenburger R and Delville J-P 2007 {\it Phys. Rev. Lett.} {\bf 98} 133601.
\bibitem{baroud07}
Baroud C N, de Saint Vincent M R and Delville J-P 2007 {\it Lab on a Chip} {\bf 7} 1029.
\bibitem{brasselet08}
Brasselet E, Wunenburger R and Delville J-P 2008 {\it Phys. Rev. Lett.} {\bf 101} 014501.
\bibitem{chraibi08}
Chra\"{\i}bi H, Lasseux D, Arquis E, Wunenburger R and Delville J-P 2008 {\it Eur.\ J.\ Mech.\ B/Fluids} {\bf 27} 419.
\bibitem{chraibi10}
Chra\"{\i}bi H, Lasseux D, Wunenburger R, Arquis E and Delville J-P 2010 {\it Eur.\ Phys.\ J.\ E} {\bf 32} 43.
\bibitem{wunenburger11}
Wunenburger R, Issenmann B, Brasselet E, Loussert C, Hourtane V and Delville J-P 2011 {\it J.\ Fluid Mech.} {\bf 666} 273.
\bibitem{hallanger05}
Hallanger A, Brevik I, Haaland S and Sollie R 2005 {\it Phys.\ Rev.\ E} {\bf 71} 056601.
\bibitem{birkeland08}
Birkeland O J and Brevik I 2008 {\it Phys. Rev. E} {\bf 78} 066314.
\bibitem{stratton41}
Stratton J A 1941 {\it Electromagnetic Theory} (New York: McGraw-Hill).
\bibitem{ellingsen11b}
Ellingsen S {\AA} and Brevik I 2011 {\it Phys.\ Fluids} {\bf 23} 096101.
\bibitem{ellingsen12b}
Ellingsen S {\AA} and Brevik I 2012 {\it Opt.\ Lett.} {\bf 37} 1928.
%\bibitem{BookAbramowitz64}
%Abramowitz M and Stegun I A 1964 {\it Handbook of Mathematical Functions} (New York: Dover).
%\bibitem{NIST10}
%Olver F W J, Lozier D W, Boisvert R F and Clark C W 2010 {\it NIST Handbook of Mathematcial Functions} (Cambridge: Cambridge University Press)
%\bibitem{bvp4c}
%Shampine L F, Reichelt M W and Kierzenka J, Solving Boundary Value Problems for Ordinary Differential Equations in MATLAB with bvp4c {\it %http://www.mathworks.com/bvp\_tutorial}.
\bibitem{barton89}
Barton J P, Alexander D R and Schaub S A 1989 {\it J.~Appl.\ Phys.} {\bf 66} 4594.
\bibitem{jackson}
Jackson J D 1999 {\it Classical Electrodynamics} 3rd ed.\ (New York: Wiley).
%New refs
\bibitem{eggers97}
Eggers J 1997 {\it Rev. Mod. Phys.} {\bf 69} 865.
\end{thebibliography}
\end{document}